\documentclass[twocolumn,journal]{IEEEtran}
\usepackage[T1]{fontenc}
\usepackage[latin9]{inputenc}
\usepackage{mathrsfs}
\usepackage{amsmath}
\usepackage{amssymb}
\usepackage{graphicx}
\usepackage[unicode=true,
 bookmarks=true,bookmarksnumbered=true,bookmarksopen=true,bookmarksopenlevel=1,
 breaklinks=false,pdfborder={0 0 0},pdfborderstyle={},backref=false,colorlinks=false]
 {hyperref}
\hypersetup{pdftitle={Your Title},
 pdfauthor={Your Name},
 pdfpagelayout=OneColumn, pdfnewwindow=true, pdfstartview=XYZ, plainpages=false}

\makeatletter

\providecommand{\tabularnewline}{\\}

\let\oldforeign@language\foreign@language
\DeclareRobustCommand{\foreign@language}[1]{%
  \lowercase{\oldforeign@language{#1}}}

\usepackage[caption=false,font=footnotesize]{subfig}
\usepackage{braket}
\usepackage[nosort]{cite}

\usepackage{tikz}
\usetikzlibrary{calc}
\newcommand{\tikzmark}[1]{\tikz[baseline,remember picture] \coordinate (#1) {};}

\usepackage{mathrsfs}

\newtheorem{theorempao}{Theorem}

\makeatother

\begin{document}
\title{The Enhanced Gaussian Noise Model extended to Polarization-dependent
Loss}
\author{Paolo Serena,~\IEEEmembership{Member,~IEEE,} Chiara Lasagni, and~Alberto
Bononi,~\IEEEmembership{Senior~Member,~IEEE}\thanks{Manuscript received xxxxxxxxx xx, xxxx; accepted xxxxxxxxx xx, xxxx.
Date of publication xxxxxxxxx xx, xxxx; date of current version xxxxxxxxx
xx, xxxx. This work was supported by Nokia Bell-labs, Villarceaux,
France. \emph{(Corresponding author: Paolo Serena.)}}\thanks{The authors are with the Department of Ingegneria e Architettura,
Universit\`a di Parma, Parma 43124, Italy (e-mail: paolo.serena@unipr.it,
chiara.lasagni@unipr.it, alberto.bononi@unipr.it).}\thanks{Color versions of one or more of the figures in this paper are available
online at http://ieeexplore.ieee.org.}\thanks{Digital Object Identifier xx.xxxx/JLT.xxxx.xxxxxxx.}}
\markboth{Journal of Lightwave Technology}{Paolo Serena \MakeLowercase{\emph{et al.}}}
\IEEEpubid{}
\maketitle
\begin{abstract}
We show how to extend the enhanced Gaussian noise (EGN) model to account
for polarization-dependent loss (PDL) of optical devices placed along
a fiber-optic link. We provide a comprehensive theory highlighting
the relationships between the time, frequency, and polarization domains
in the presence of fiber nonlinear Kerr effect and amplified spontaneous
emission. We double-check the new model with split-step Fourier method
(SSFM) simulations showing very good accuracy. The model can be efficiently
exploited to estimate low values of outage probabilities induced by
PDL with computational times orders of magnitude faster than the SSFM,
thus opening new opportunities in the design of optical communication
links.
\end{abstract}


\IEEEpeerreviewmaketitle{}

\section{Introduction}

\IEEEPARstart{P}{olarization}-dependent loss (PDL) expresses the
dependence of the loss of an optical device on the state of polarization
of the input electromagnetic field \cite{Damask}. PDL induces crosstalk
between the polarization tributaries and an unequal loss of energy,
which are particularly detrimental in polarization-division multiplexing
(PDM) transmissions.

Although typical optical fibers show negligible PDL, PDL may be relevant
in optical devices, such as the Erbium-doped fiber amplifiers (EDFA)
and inside the wavelength selective switches (WSS) of reconfigurable
optical add/drop multiplexers (ROADM) \cite{Nelson}.

The axes of maximum/minimum PDL fluctuate randomly over times much
longer than the coded data block duration, thus making the optical
channel stochastic and non-ergodic. Hence the analysis with PDL should
not focus on the average performance, such as the average signal-to-noise
ratio (SNR), but rather on the statistics of the SNR. Of particular
concern is the outage probability, i.e., the probability that the
SNR falls below a given threshold. Because of the random fluctuations,
the problem of estimating such as probability is particularly challenging,
especially in numerical simulations where low outage values call for
many time-consuming simulations, but also in experiments where collecting
many observations may require a huge amount of resources to save and
post-process the results. 

Such difficulties stimulated the development of theoretical models
for quick estimation of the PDL effects. Most of the literature focused
on the interplay between PDL and amplified spontaneous emission (ASE)
noise in the linear regime. Remarkable results have been provided
by Gisin \cite{Gisin}, who found the statistics of the resulting
PDL after concatenation of many devices, and by Mecozzi and Shtaif
\cite{Mecozzi_PTL_PDL} that investigated the asymptotic properties
of PDL showing its Maxwellian statistics when expressed in dB. The
implications of the interplay PDL-ASE on the SNR have been investigated
by the same authors in \cite{Mecozzi_JLT_04} and by Shtaif in \cite{Shtaif}.
The implications of PDL on the channel capacity has been investigated
by Nafta \emph{et al.} in \cite{Nafta}. A quaternion approach to
analytically investigate PDL has been proposed by Karlsson and Petersson
in \cite{Karlsson}.

The interplay between PDL and the fiber nonlinear Kerr effect received
much less analytical attention, and most of the literature focused
on numerical/experimental investigations \cite{Tao,Xie,Vassilieva_OFC,Ivan_Amir,Chin,Rossi,Cartledge_JLT19,Cartledge_PDL_II}.
Such investigations showed contrasting results, since PDL showed limited
interaction with the Kerr effect in \cite{Chin} while a non-negligible
interaction has been pointed out, for instance, in \cite{Vassilieva_OFC,Rossi}. 

In modern optical communication systems, it is customary to analyze
the performance of the link by employing perturbative models because
of their simplicity. Among the available models in the literature,
particular attention has been captured by the Gaussian noise (GN)
model \cite{Poggiolini_GN} and its advanced version, the enhanced
Gaussian noise (EGN) model \cite{Carena_EGN,Serena_EGN}, also referred
to as nonlinear interference noise (NLIN) model \cite{Dar_NLIN}.
Such models showed excellent accuracy in a wide range of optical links,
with savings in computational time of more than an order of magnitude
compared with traditional models, such as the split-step Fourier method
(SSFM).

We extended the scalar theory of the GN model by including polarization
effects in \cite{serena_GN_JLT}, and first included PDL in the GN
model framework in \cite{Serena_PDL_ecoc18} for a quick estimation
of the probability density function (PDF) of the nonlinear interference
(NLI).

In this work, besides providing a novel mathematical formalism to
cope with PDL in the GN model, we show how to account for PDL even
in the EGN model. The general theory will be double-checked against
SSFM simulations. 

The advantages of using the new model as well as some interesting
implications, such as the scaling of the outage probability with power,
will be discussed.

The paper is organized as follows: in Section~\ref{sec:PDL-extended-EGN-model}
we show the main theory, based on some results provided in the Appendices;
in Section~\ref{sec:Numerical-validation} we validate the model.
Finally, in Section~\ref{sec:Conclusions} we draw our main conclusions.

\section{PDL-extended EGN model\label{sec:PDL-extended-EGN-model}}

We adopt the following bra-ket notation 
\[
\ket{\tilde{A}(\omega)}\triangleq\left[\begin{array}{c}
\tilde{A}_{x}(\omega)\\
\tilde{A}_{y}(\omega)
\end{array}\right],\quad\bra{\tilde{A}(\omega)}\triangleq\left[\tilde{A}_{x}^{*}(\omega),~\tilde{A}_{y}^{*}(\omega)\right]
\]
where $\tilde{A}_{x,y}(\omega)$ indicate the Fourier transform of
the two polarization tributaries of the transmitted electric field,
with $\omega$ the angular frequency. 

Under a first-order perturbative approximation, the received electric
field $\ket{\tilde{A}_{\text{R}}(\omega)}$ can be related to the
transmitted one $\ket{\tilde{A}(\omega)}$ by \cite{Mecozzi}:
\begin{equation}
\ket{\tilde{A}_{\text{R}}(\omega)}\approx\mathbf{T}(z,\omega)\left(\ket{\tilde{A}(\omega)}+\ket{\tilde{w}(\omega)}+\ket{\tilde{n}(\omega)}\right)\label{eq:RP1}
\end{equation}
with $\ket{\tilde{w}(\omega)}$ and $\ket{\tilde{n}(\omega)}$ the
ASE noise and the signal NLI, respectively, and $\mathbf{T}(z,\omega)$
a $2\times2$ matrix accounting for all linear impairments from input
to coordinate $z$. Such a matrix can be separated into a scalar and
a polarization-dependent contribution:
\begin{align}
\mathbf{T}(z,\omega) & =e^{\vartheta(z,\omega)}\mathbf{U}(z)\nonumber \\
\vartheta(z,\omega) & \triangleq-\int_{0}^{z}\left(\frac{\alpha(\xi)}{2}+j\beta(\xi,\omega)\right)\text{d}\xi\label{eq:lineareff}
\end{align}
with $\alpha$ the fiber attenuation and $\beta$ the imaginary part
of the propagation constant. The matrix $\mathbf{U}$ accounts for
a frequency-independent PDL accumulated up to coordinate $z$. We
assume lumped PDL (e.g., WSS and EDFA) at coordinates $z_{p}:~p=0,\ldots,N-1$,
with $z_{0}=0$. The matrix $\mathbf{U}$ depends on the $k$th device
at coordinate $z_{k}\le z$ with PDL matrix $\mathbf{M}_{k}$ by \cite{Damask}:
\begin{align}
\mathbf{U}(z) & =\mathbf{M}_{p}\mathbf{M}_{p-1}\cdots\mathbf{M}_{0},\qquad z_{p}<z<z_{p+1}\nonumber \\
\mathbf{M}_{k} & \triangleq\mathbf{W}_{k}^{\dagger}\left[\begin{array}{cc}
\sqrt{1+\Gamma_{k}} & 0\\
0 & \sqrt{1-\Gamma_{k}}
\end{array}\right]\mathbf{W}_{k}\label{eq:Mk}
\end{align}
where $\dagger$ indicates transpose-conjugate, the $\mathbf{W}_{k}$
are matrices uniformly distributed in the set of the $2\times2$ unitary
random matrices (Haar matrices), while $\Gamma_{k}$ defines the PDL
$\rho_{k}$ by $\rho_{k}\triangleq\left(1+\Gamma_{k}\right)/\left(1-\Gamma_{k}\right)$.
The PDL is usually expressed in dB by $20\log_{10}(\rho_{k})$. 

In the case of many identically distributed PDL elements, the resulting
PDL of the link, expressed in dB, follows a Maxwellian distribution
\cite{Mecozzi_PTL_PDL} with an average value scaling with $\sqrt{N}$.

\begin{figure}[tbh]
\begin{centering}
\includegraphics[width=1\linewidth]{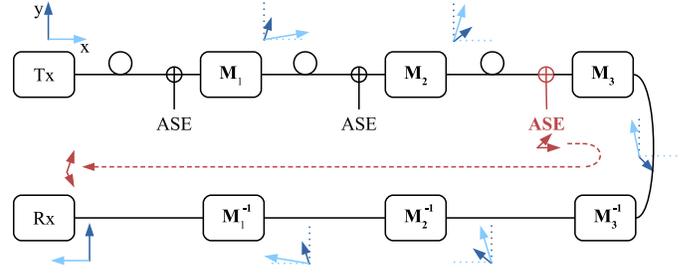}
\par\end{centering}
\caption{\label{fig:schema}Sketch of the effects experienced by the signal
$\ket{A}$ and the ASE $\ket{w}$ along propagation. The zero-forcing
equalization assumption is equivalent to a round-trip propagation
of the signal, thus returning equal to itself at reception. However,
ASE follows an incomplete round-trip, resulting in PDL. }
\end{figure}

Matrix $\mathbf{W}_{k}$ is statistically independent of matrix $\mathbf{W}_{n}$,
with $k\ne n$. Such matrices are slowly varying in time compared
to the symbol timing, hence while each polarization tributary preserves
its average power while crossing the generic PDL element, each PDM
data-block experiences a power unbalance between polarization tributaries.
Such an imbalance can be removed at the receiver side by performing
linear equalization, for instance by the zero-forcing equalizer $\mathbf{T}^{-1}(z,\omega)$.
However, PDL remains both in the ASE and the NLI. Fig.~\ref{fig:schema}
sketches the idea for ASE. In this work, we assume $\mathbf{W}_{k}$
a random variable, thus time-independent. 

The received ASE $\ket{\tilde{w}(\omega)}$ is related to the ASE
$\ket{\tilde{w}_{m}(\omega)}$ outgoing the in-line amplifier at coordinate
$z_{m}$ by:
\begin{equation}
\ket{\tilde{w}(\omega)}=\sum_{m=1}^{M}\mathbf{M}_{0}^{-1}\mathbf{M}_{1}^{-1}\cdots\mathbf{M}_{m-1}^{-1}\ket{\tilde{w}_{m}(\omega)}~.\label{eq:ASE_prodsum}
\end{equation}
 with $M$ the number of amplifiers. For instance, concerning Fig.~\ref{fig:schema}
referred to an ideal source, i.e.,\textbf{ $\mathbf{M}_{0}=\mathbf{I}$}
with $\mathbf{I}$ the identity matrix, we have:\textbf{ }
\begin{align}
 & \ket{\tilde{w}(\omega)}=\ket{\tilde{w}_{1}(\omega)}+\mathbf{M}_{1}^{-1}\ket{\tilde{w}_{2}(\omega)}+\mathbf{M}_{1}^{-1}\mathbf{M}_{2}^{-1}\ket{\tilde{w}_{3}(\omega)}\nonumber \\
 & =\ket{\tilde{w}_{1}(\omega)}+\mathbf{U}(z_{1})^{-1}\ket{\tilde{w}_{2}(\omega)}+\mathbf{U}(z_{2})^{-1}\ket{\tilde{w}_{3}(\omega)}~.\label{eq:ASE-equal}
\end{align}
In Appendix~\ref{App:Nonlinear-interference} we show that the NLI
takes the expression:
\begin{multline}
\ket{\tilde{n}(\omega)}=-j\iint_{-\infty}^{\infty}\sum_{p=0}^{N-1}\eta_{p}(\omega,\omega_{1},\omega_{2})\times\\
\braket{\tilde{A}(\omega+\omega_{1}+\omega_{2})|\mathbf{P}(z_{p})|\tilde{A}(\omega+\omega_{2})}\ket{\tilde{A}(\omega+\omega_{1})}\frac{\text{d}\omega_{1}}{2\pi}\frac{\text{d}\omega_{2}}{2\pi}\label{eq:aw}
\end{multline}
with $\eta_{p}$ the kernel of the optical fiber in the segment $[z_{p},~z_{p+1}]$,
and $\mathbf{P}(z_{p})\triangleq\mathbf{U}^{\dagger}(z_{p})\mathbf{U}(z_{p})$.
Please note that, without PDL, $\mathbf{P}=\mathbf{I}$. Eq. (\ref{eq:aw})
clearly shows the four-wave mixing (FWM) process generating the NLI. 

Both the NLI and the ASE accumulate linearly with the propagation
distance, both being additive under the model assumptions. Moreover,
by comparing (\ref{eq:ASE_prodsum}) and (\ref{eq:aw}), we observe
that after zero-forcing equalization they both depend on the PDL accumulated
\emph{before} their generation. However, such a dependence follows
completely different relationships. In particular, the quadratic dependence
on the entries of matrix $\mathbf{U}(z_{p})$ in the NLI is expected
to induce larger random fluctuations of the SNR compared to the linear
dependence in the ASE case. 

We now introduce a discrete-time channel model relating the transmitted/received
data symbol \cite{Mecozzi}. We assume $\ket{A(t)}$ is a wavelength
division multiplex (WDM) of linearly modulated digital signals:
\begin{equation}
\ket{A(t)}=\sum_{n=-\infty}^{\infty}\sum_{h=1}^{{\scriptscriptstyle \text{\# channels}}}\sum_{k\in(x,y)}a_{nhk}p_{h}(t-nT_{h})e^{j\Omega_{h}t}\ket{k}\label{eq:PAM_orig}
\end{equation}
where $a_{nhk}$ is the digital symbol (e.g., quadrature amplitude
modulation, QAM) at time $n$, WDM channel-index $h$, and polarization
$k$; $p_{h}(t-nT_{h})$ is the supporting pulse at the $n$th symbol
time of duration $T_{h}$ and modulated at carrier frequency $\Omega_{h}$.
We find it useful to compact the notation by calling $a_{\mathbf{n}}$
the generic (scalar) information symbol per \emph{(time, frequency,
space) }channel use, where the vector $\mathbf{n}$ should be read
as:
\begin{align}
\nonumber \\
\mathbf{n} & =\left[\tikzmark{time}n_{1},~\tikzmark{freq}n_{2},~\tikzmark{space}n_{3}\right]~.\label{eq:vectorindex}\\
\nonumber 
\end{align}
\begin{tikzpicture}[remember picture,overlay] 
\draw[<-]    ([shift={(4pt,-2pt)}]time) |- ([shift={(-10pt,-8pt)}]time)    node[anchor=east] {$\scriptstyle \text{time}$};  
\draw[<-]    ([shift={(4pt,-2pt)}]freq) |- ([shift={(14pt,-8pt)}]freq)    node[anchor=west] {$\scriptstyle \text{frequency}$}; 
\draw[<-]    ([shift={(4pt,6pt)}]space) |- ([shift={(14pt,12pt)}]space)    node[anchor=west] {$\scriptstyle \text{space}$}; 
\end{tikzpicture}We will refer to $a_{\mathbf{n}}$ as an \emph{atom}\textbf{\emph{
}}of the source information.\textbf{ }This way, by adopting signal-space
representation concepts, eq. (\ref{eq:PAM_orig}) in the Fourier domain
can be written as:
\begin{equation}
\ket{\tilde{A}(\omega)}=\sum_{\mathbf{n}}a_{\mathbf{n}}\ket{\tilde{G}_{\mathbf{n}}(\omega)}\label{eq:atom}
\end{equation}
where $\sum_{\mathbf{n}}$ stands for all summations in (\ref{eq:PAM_orig})\emph{
}and the basis functions are:
\[
\ket{\tilde{G}_{\mathbf{n}}(\omega)}\triangleq\tilde{p}_{n_{2}}(\omega-\Omega_{n_{2}})e^{-j(\omega-\Omega_{n_{2}})n_{1}T_{n_{2}}}\ket{n_{3}}~.
\]

We assume the detector performs demodulation, matched filtering, sampling
and average carrier phase recovery. In our framework, the first three
operations correspond to the inner product $\int_{-\infty}^{\infty}\braket{\tilde{G}_{\mathbf{i}}(\omega)|\tilde{A}(\omega)}\frac{\text{d}\omega}{2\pi}$.
In particular, such operation results in $a_{\mathbf{i}}$ in absence
of impairments when using orthonormal basis functions, such as root-raised
cosine pulses with non-overlapping spectrum among channels:
\begin{equation}
\sum_{\mathbf{n}}a_{\mathbf{n}}\int_{-\infty}^{\infty}\braket{\tilde{G}_{\mathbf{i}}(\omega)|\tilde{G}_{\mathbf{n}}(\omega)}\frac{\text{d}\omega}{2\pi}=a_{\mathbf{i}}~.\label{eq:ai}
\end{equation}

By following similar steps as \cite{Mecozzi,Dar_first}, from (\ref{eq:RP1})
we get the following discrete-time channel model relating the transmitted
atom $a_{\mathbf{i}}$ to the received one $u_{\mathbf{i}}$:
\[
u_{\mathbf{i}}=a_{\mathbf{i}}+w_{\mathbf{i}}+n_{\mathbf{i}}
\]
where $w_{\mathbf{i}}$ and $n_{\mathbf{i}}$ are the sampled ASE
and NLI, respectively: 
\begin{align}
w_{\mathbf{i}} & =\int_{-\infty}^{\infty}\braket{\tilde{G}_{\mathbf{i}}(\omega)|\tilde{w}(\omega)}\frac{\text{d}\omega}{2\pi}\nonumber \\
n_{\mathbf{i}} & =-j\sum_{\mathbf{k},\mathbf{m},\mathbf{n}}a_{\mathbf{k}}^{*}a_{\mathbf{m}}a_{\mathbf{n}}{\cal X}_{\mathbf{kmni}}~.\label{eq:DTM-1}
\end{align}
${\cal X}_{\mathbf{kmni}}$ is a tensor weighting the four-atom mixing
(FAM) at the symbol level:
\begin{align}
{\cal X}_{\mathbf{kmni}} & =\sum_{p=0}^{N-1}\iiint_{-\infty}^{\infty}\eta_{p}(\omega,\omega_{1},\omega_{2})\nonumber \\
 & \times\braket{\tilde{G}_{\mathbf{k}}(\omega+\omega_{1}+\omega_{2})|\mathbf{P}(z_{p})|\tilde{G}_{\mathbf{m}}(\omega+\omega_{2})}\nonumber \\
 & \times\braket{\tilde{G}_{\mathbf{i}}(\omega)|\tilde{G}_{\mathbf{n}}(\omega+\omega_{1})}\frac{\text{d}\omega_{1}}{2\pi}\frac{\text{d}\omega_{2}}{2\pi}\frac{\text{d}\omega}{2\pi}\nonumber \\
 & =\sum_{p=0}^{N-1}P_{k_{3}m_{3}}(z_{p})\delta_{i_{3}n_{3}}{\cal S}_{\mathbf{kmni}}(z_{p})\label{eq:tensor_chi}
\end{align}
where the $\delta$ indicates Kronecker's delta, and in the final
identity of (\ref{eq:tensor_chi}) we expanded the tensor in terms
of the tensor ${\cal S}_{\mathbf{kmni}}$ weighting the FWM interaction
at the scalar level:
\begin{multline}
{\cal S}_{\mathbf{kmni}}(z_{p})\triangleq\iiint_{-\infty}^{\infty}\eta_{p}(\omega,\omega_{1},\omega_{2})\times\\
\tilde{G}_{\mathbf{k}}^{*}(\omega+\omega_{1}+\omega_{2})\tilde{G}_{\mathbf{m}}(\omega+\omega_{2})\tilde{G}_{\mathbf{i}}^{*}(\omega)\tilde{G}_{\mathbf{n}}(\omega+\omega_{1})\frac{\text{d}\omega_{1}}{2\pi}\frac{\text{d}\omega_{2}}{2\pi}\frac{\text{d}\omega}{2\pi}\label{eq:tensor_S}
\end{multline}
where $\tilde{G}_{\mathbf{n}}(\omega)$ is defined in implicit form
by $\ket{\tilde{G}_{\mathbf{n}}(\omega)}\triangleq\tilde{G}_{\mathbf{n}}(\omega)\ket{n_{3}}$.
It is worth noting that in the special, yet relevant, case of a homogeneous
link in absence of PDL, the summation $\sum_{p}$ in (\ref{eq:tensor_chi})
can be closed with some advantage for numerical purposes and simplicity.

We now evaluate the covariance of ASE and NLI atoms when acting alone.

\subsection{ASE variance}

The PDL impact on ASE has been investigated in several papers in the
literature \cite{Karlsson,Shtaif,Tao,Damask}, whose main results
we now rephrase in our notation.

By definition, the variance of the ASE atom $\mathbf{i}$ is $\sigma_{{\scriptscriptstyle \text{ASE}}}^{2}=\mathbb{E}[w_{\mathbf{i}}w_{\mathbf{i}}^{*}]$,
with $\mathbb{E}$ indicating expectation. We evaluate it focusing
on a link with independent and identically distributed ASE sources,
hence with \cite[p. 418]{Papoulis}:
\begin{equation}
\mathbb{E}\left[\ket{\tilde{w}_{m}(\omega)}\bra{\tilde{w}_{p}(\mu)}\right]=\frac{N_{0}}{2}\delta(\omega-\mu)\delta_{mp}\mathbf{I}\label{eq:ASE_iid}
\end{equation}
where the two $\delta$ indicate Dirac/Kronecker's delta, respectively,
while $N_{0}$ is the one-sided, dual-polarization, power spectral
density (PSD) of ASE per amplifier. $N_{0}$ is related to the noise
figure $F$ and the gain $G$ by $N_{0}=h\nu FG$, with $h$ Planck's
constant and $\nu$ carrier frequency.

Let $B$ be the noise equivalent bandwidth of the receiver. By using
(\ref{eq:ASE_prodsum}) and the orthogonality property (\ref{eq:ai})
in (\ref{eq:ASE_iid}) we have:
\begin{equation}
\sigma_{{\scriptscriptstyle \text{ASE}}}^{2}=\frac{N_{0}B}{2}\sum_{p=1}^{M}P_{i_{3}i_{3}}^{-1}(z_{p})\label{eq:rhoij_ASE}
\end{equation}
with $M$ the number of optical amplifiers in the link. Please note
that the matrices $\mathbf{P}(z_{p})$ and hence their elements $P_{i_{3}j_{3}}(z_{p})$
are not independent but related by the concatenation rule of PDL \cite{Karlsson}. 

With PDL it is more interesting to deal with the SNR per polarization,
because of the asymmetrical behavior of noise power. With reference
to the generic polarization $i$:
\begin{equation}
\text{SNR}_{{\scriptscriptstyle \text{ASE}}}^{i}=\frac{S_{i}}{\sigma_{{\scriptscriptstyle \text{ASE}}}^{2}}=\frac{\text{SNR}_{{\scriptscriptstyle \text{ASE}}}^{i}(\text{PDL=0})}{\frac{1}{M}\sum_{p=1}^{M}P_{ii}^{-1}(z_{p})},\qquad i\in(x,y)\label{eq:SNR_ASE}
\end{equation}
where $S_{i}$ is the signal power on polarization $i$ and $\text{SNR}_{{\scriptscriptstyle \text{ASE}}}^{i}(\text{PDL=0})=S_{i}/\left(\frac{N_{0}MB}{2}\right)$.
The denominator of (\ref{eq:SNR_ASE}), equal to the span-average
of $P_{ii}^{-1}$, is the random PDL loss/gain per polarization.

\subsection{NLI variance}

The FWM process underpinning the NLI is formally identical to the
scalar case, with just a different weighting tensor. However, some
symmetries cease to hold, hence the master theorem at the heart of
the EGN model (see Appendix~\ref{App:Master-theorem}) must be properly
generalized by taking care of such a novelty. We observe the following
symmetries in indexing:
\begin{align}
{\cal X}_{\mathbf{\mathbf{kmni}}} & ={\cal X}_{\mathbf{m\mathbf{ki}n}}^{*},\qquad(\text{always})\nonumber \\
{\cal X}_{\mathbf{kmni}} & ={\cal X}_{\mathbf{knmi}},\qquad(\text{no PDL})~.\label{eq:X_symmetry}
\end{align}
The breakdown of the last symmetry induces a small modification in
the master theorem as detailed in Appendix~\ref{App:Master-theorem}.

With such ingredients, after carrier phase estimation the variance
of the NLI atom $n_{\mathbf{i}}^{\prime}$, $\sigma_{{\scriptscriptstyle \text{NLI}}}^{2}=\mathbb{E}[n_{\mathbf{i}}^{\prime}n_{\mathbf{i}}^{\prime*}]$,
can be found by plugging (\ref{eq:tensor_chi}) into the master theorem
(\ref{eq:Main_thm}). Such a variance can be split into the GN, fourth-order
noise (FON) \cite{Mecozzi}, and higher-order noise (HON) \cite{Carena_EGN,Serena_EGN,Dar_NLIN}
contributions:
\[
\sigma_{{\scriptscriptstyle \text{NLI}}}^{2}=\sigma_{{\scriptscriptstyle \text{GN}}}^{2}+\underbrace{\sigma_{{\scriptscriptstyle \text{FON}}}^{2}+\sigma_{{\scriptscriptstyle \text{HON}}}^{2}}_{{\scriptscriptstyle \text{EGN correction}}}~.
\]
We now analyze the contributions for independent and identically distributed
data symbols. The result depends on the statistical cumulants of the
symbols \cite{Serena_EGN}, which are related to the main moments
$\mu_{n}\triangleq\mathbb{E}[|a_{\mathbf{k}}|^{n}]$ by:
\begin{align*}
\kappa_{1} & =\mu_{2}\\
\kappa_{2} & =\mu_{4}-2\mu_{2}^{2}\\
\kappa_{3} & =\mu_{6}-9\mu_{4}\mu_{2}+12\mu_{2}^{3}~.
\end{align*}

\subsubsection{GN term}

As outlined in Appendix~\ref{App:Master-theorem}, the GN contribution
to the variance of polarization $i_{3}\in(x,y)$ is:

\begin{multline*}
\sigma_{{\scriptscriptstyle \text{GN}}}^{2}=\kappa_{1}^{3}\sum_{\mathbf{k,m,n}}{\cal X}_{\mathbf{kmni}}\left({\cal X}_{\mathbf{kmni}}^{*}+{\cal X}_{\mathbf{knmi}}^{*}\right)\\
=\kappa_{1}^{3}\sum_{\mathbf{k,m,n}}\delta_{i_{3}n_{3}}\sum_{p,\ell=0}^{N-1}{\cal S}_{\mathbf{kmni}}(z_{p}){\cal S}_{\mathbf{kmni}}^{*}(z_{\ell})\\
\times\Big(P_{k_{3}m_{3}}(z_{p})P_{k_{3}m_{3}}^{*}(z_{\ell})+\delta_{i_{3}m_{3}}P_{k_{3}m_{3}}(z_{p})P_{k_{3}n_{3}}^{*}(z_{\ell})\Big)
\end{multline*}
where we used (\ref{eq:tensor_chi}). Such a result can be easily
generalized to the spatial-covariance matrix. We introduce the $2\times2$
GN covariance matrix between the polarizations,\textbf{ $\mathbf{K}_{{\scriptscriptstyle \text{GN}}}:K_{i_{3}j_{3}}^{{\scriptscriptstyle \text{(GN)}}}=\mathbb{E}[n_{\mathbf{i}}^{'}n_{\mathbf{j}}^{'*}],~i_{1,2}=j_{1,2},~(i_{3},j_{3})\in(x,y)$,}
which takes the elegant form \cite{Serena_PDL_ecoc18}:
\begin{multline}
\mathbf{K}_{{\scriptscriptstyle \text{GN}}}\triangleq\left[\begin{array}{cc}
\text{var}\left(\text{NLI}_{x}^{{\scriptscriptstyle (\text{GN})}}\right) & \text{cov}\left(\text{NLI}_{x}^{{\scriptscriptstyle (\text{GN})}},\text{NLI}_{y}^{{\scriptscriptstyle (\text{GN})}}\right)\\
\text{cov}\left(\text{NLI}_{x}^{{\scriptscriptstyle (\text{GN})}},\text{NLI}_{y}^{{\scriptscriptstyle (\text{GN})}}\right) & \text{var}\left(\text{NLI}_{y}^{{\scriptscriptstyle (\text{GN})}}\right)
\end{array}\right]\\
=\sum_{p,\ell=0}^{N-1}\rho_{{\scriptscriptstyle \text{GN}}}(p,\ell)\Big(\mathrm{Tr}\left[\mathbf{P}(z_{p})\mathbf{P}^{\dagger}(z_{\ell})\right]\mathbf{I}+\mathbf{P}(z_{p})\mathbf{P}^{\dagger}(z_{\ell})\Big)\label{eq:covarianceK}
\end{multline}
where $\rho_{{\scriptscriptstyle \text{GN}}}(p,\ell)\triangleq\sum{\cal S}_{\mathbf{kmni}}(z_{p}){\cal S}_{\mathbf{kmni}}^{*}(z_{\ell})$,
with the summations limited to the temporal and frequency indexes,
is the \emph{scalar} cross-correlation between the NLI accumulated
in trunk $p$ and trunk $\ell$, while Tr indicates the trace of a
matrix. It is worth noting that in absence of PDL we have \cite{Mecozzi,Dar_NLIN}:
\begin{multline}
\!\!\!\!\!\sigma_{{\scriptscriptstyle \text{GN}}}^{2}(\text{no PDL})=\!\sum_{p,\ell=0}^{N-1}\rho_{{\scriptscriptstyle \text{GN}}}(p,\ell)=\!\!\sum_{h,r,s=1}^{{\scriptscriptstyle \text{\# channels}}}\iiint_{-\infty}^{\infty}\!|\eta(\omega,\omega_{1},\omega_{2})|^{2}\\
\times\left|\tilde{P}_{i}^{*}(\omega-\Omega_{i})\right|^{2}\left|\tilde{P}_{h}^{*}(\omega+\omega_{1}+\omega_{2}-\Omega_{h})\right|^{2}\\
\times\left|\tilde{P}_{s}(\omega+\omega_{2}-\Omega_{s})\right|^{2}\left|\tilde{P}_{r}(\omega+\omega_{1}-\Omega_{r})\right|^{2}\frac{\text{d}\omega}{2\pi}\frac{\text{d}\omega_{1}}{2\pi}\frac{\text{d}\omega_{2}}{2\pi}~.\label{eq:GN_classic}
\end{multline}
where the term $\eta$ is the fiber-kernel of the entire link, see
Appendix~\ref{App:Nonlinear-interference}. We also observe that,
since without PDL $\mathbf{P}(z_{p})=\mathbf{I}$ for each $p$, it
is $\mathbf{K}_{{\scriptscriptstyle \text{GN}}}=3\mathbf{I}$ \cite{Serena_EGN}.

The right-hand side in (\ref{eq:GN_classic}) is well known in the
literature, see, e.g., \cite{Mecozzi}. The main contribution of this
work is that eq. (\ref{eq:covarianceK}) generalizes the scalar result
(\ref{eq:GN_classic}) to the case with PDL. We note that now we need
to know all trunk cross-correlations, while in the scalar case such
information was not required. However, the matrix of elements $\rho_{{\scriptscriptstyle \text{GN}}}(p,\ell)$
is a Toeplitz matrix, hence it can be calculated in a short simulation
(preload), usually of the order of seconds, for instance with the
algorithm \cite{Dar_NLIN}. Once $\rho_{{\scriptscriptstyle \text{GN}}}$
is available, the PDL statistics can be evaluated very quickly by
computing the matrix in (\ref{eq:covarianceK}).

The SNR of the generic polarization $i$ finally is \cite{Poggiolini_GN}:
\begin{equation}
\text{SNR}_{{\scriptscriptstyle \text{GN}}}^{i}=\frac{1}{K_{{\scriptscriptstyle ii}}^{{\scriptscriptstyle \text{(GN)}}}S_{i}^{2}},\qquad i\in(x,y)\label{eq:SNRgn}
\end{equation}
which, contrary to ASE, cannot be expressed in closed-form in terms
of the SNR without PDL.

\subsubsection{FON terms}

The FON variance of polarization $i_{3}\in(x,y)$ is derived in Appendix~\ref{App:Master-theorem},
here repeated for convenience:
\begin{multline}
\sigma_{{\scriptscriptstyle \text{FON}}}^{2}=\kappa_{2}\kappa_{1}\sum_{\mathbf{k,n}}\Big(\left|{\cal X}_{\mathbf{kkni}}+{\cal X}_{\mathbf{knki}}\right|^{2}+\left|{\cal X}_{\mathbf{nkki}}\right|^{2}\Big)~.\label{eq:FON}
\end{multline}
In Appendix~\ref{App:Master-theorem} we label the first absolute
value in (\ref{eq:FON}) by F4 and the second by Q4. With similar
steps done for the GN counterpart, we introduce the FON covariance
matrix $\mathbf{K}_{{\scriptscriptstyle \text{FON}}}$ between spatial
coordinates. After inserting (\ref{eq:tensor_chi}) in (\ref{eq:FON})
we obtain:
\[
\mathbf{K}_{{\scriptscriptstyle \text{FON}}}=\sum_{p,\ell=0}^{N-1}\Big(\rho_{{\scriptscriptstyle \text{F4}}}(p,\ell)\mathbf{F}(z_{p},z_{\ell})+\rho_{{\scriptscriptstyle \text{Q4}}}(p,\ell)\mathbf{Q}(z_{p},z_{\ell})\Big)
\]
where $\rho_{{\scriptscriptstyle \text{F4}}}(p,\ell)\triangleq\sum{\cal S}_{\mathbf{kkni}}(z_{p}){\cal S}_{\mathbf{kkni}}^{*}(z_{\ell})$
and $\rho_{{\scriptscriptstyle \text{Q4}}}(p,\ell)\triangleq\sum{\cal S}_{\mathbf{nkki}}(z_{p}){\cal S}_{\mathbf{nkki}}^{*}(z_{\ell})$,
with the summations limited to the temporal and frequency indexes,
are the scalar cross-correlations between the two kinds of FON (see
Appendix~\ref{App:Master-theorem}) accumulated in trunk $p$ and
trunk $\ell$. They can be evaluated with the scalar EGN model \cite{Dar_NLIN,Carena_EGN,Serena_EGN}. 

Matrices $\mathbf{F}$ and $\mathbf{Q}$ are the novelty introduced
by PDL. They have the following entries:
\begin{align*}
F_{11} & =4P_{11}(z_{p})P_{11}^{*}(z_{\ell})+P_{22}(z_{p})P_{22}^{*}(z_{\ell})+P_{12}(z_{p})P_{12}^{*}(z_{\ell})\\
F_{12} & =P_{22}(z_{p})P_{21}^{*}(z_{\ell})+P_{12}(z_{p})P_{11}^{*}(z_{\ell})\\
F_{21} & =P_{11}(z_{p})P_{12}^{*}(z_{\ell})+P_{21}(z_{p})P_{22}^{*}(z_{\ell})\\
F_{22} & =4P_{22}(z_{p})P_{22}^{*}(z_{\ell})+P_{11}(z_{p})P_{11}^{*}(z_{\ell})+P_{21}(z_{p})P_{21}^{*}(z_{\ell})
\end{align*}
while matrix $\mathbf{Q}$:
\begin{align*}
Q_{11} & =P_{11}(z_{p})P_{11}^{*}(z_{\ell})+P_{21}(z_{p})P_{21}^{*}(z_{\ell})\\
Q_{12} & =Q_{21}=0\\
Q_{22} & =P_{22}(z_{p})P_{22}^{*}(z_{\ell})+P_{12}(z_{p})P_{12}^{*}(z_{\ell})~.
\end{align*}
Please note that in absence of PDL we have $\mathbf{F}=5\mathbf{I}$
and $\mathbf{Q}=\mathbf{I}$, as expected \cite{Serena_EGN}. It is
worth noting that in most of the optical links the dominant FON term
is the one related to the F4 term.

\subsubsection{HON term}

The HON is a sixth order term, labeled by Q6 in Appendix~\ref{App:Master-theorem}.
Such a term occurs only when all the six atoms joining the NLI covariance
are identical. We have:
\[
\sigma_{{\scriptscriptstyle \text{HON}}}^{2}=\kappa_{3}\sum_{\mathbf{n}}\left|{\cal X}_{\mathbf{nnni}}\right|^{2}~.
\]
We introduce a HON covariance matrix $\mathbf{K}_{{\scriptscriptstyle \text{HON}}}$
between spatial coordinates obtaining:
\[
\mathbf{K}_{{\scriptscriptstyle \text{HON}}}=\sum_{p,\ell=0}^{N-1}\rho_{{\scriptscriptstyle \text{Q6}}}(p,\ell)\mathbf{H}(z_{p},z_{\ell})
\]
with $\rho_{{\scriptscriptstyle \text{Q6}}}$ the scalar HON cross-correlation
between trunk $p$ and trunk $\ell$, while matrix $\mathbf{H}$ has
entries:
\begin{align*}
H_{ii} & =P_{ii}(z_{p})P_{ii}^{*}(z_{\ell}),\quad i=1,2\\
H_{ij} & =0,\quad i\ne j~.
\end{align*}
Please note that in absence of PDL we have $\mathbf{H}=\mathbf{I}$.
In most of the optical links, the HON term is dominated by the FON
term.

Finally, the SNR associated to the NLI is:
\[
\text{SNR}_{{\scriptscriptstyle \text{NLI}}}^{i}=\frac{1}{\left(K_{{\scriptscriptstyle ii}}^{{\scriptscriptstyle \text{(GN)}}}+K_{{\scriptscriptstyle ii}}^{{\scriptscriptstyle \text{(FON)}}}+K_{{\scriptscriptstyle ii}}^{{\scriptscriptstyle \text{(HON)}}}\right)S_{i}^{2}},\quad i\in(x,y)
\]
while the overall SNR, by neglecting ASE-NLI interaction, follows
the usual concatenation rule:
\begin{equation}
\frac{1}{\text{SNR}^{i}}=\frac{1}{\text{SNR}_{{\scriptscriptstyle \text{ASE}}}^{i}}+\frac{1}{\text{SNR}_{{\scriptscriptstyle \text{NLI}}}^{i}}~.\label{eq:SNR_total}
\end{equation}

\section{Numerical validation\label{sec:Numerical-validation}}

We checked the proposed model against SSFM based simulations. Common
parameters to all simulations are the pulse types, i.e., root-raised
cosine pulses with roll-off 0.01 sent at 49 Gbaud, the channel spacing,
50 GHz, and the optical fibers, i.e., single-mode fibers (SMF) having
length $100$ km, dispersion $D=17$ ps/nm/km, attenuation $\alpha=0.2$
dB/km, nonlinear coefficient $\gamma=1.26$ 1/W/km. The channel under
test (CUT) is in any case the central channel of the WDM.

Since the inclusion of PDL in the NLI is the main novelty of this
work, in a first test we focused on the above system without ASE.
The WDM was made of $11\times50$ GHz channels, PDM-modulated with
Gaussian distributed symbols, i.e., the capacity-achieving modulation
format for the additive white Gaussian noise (AWGN) channel. In this
case, the EGN model degenerates into the GN model \cite{Serena_PDL_ecoc18}.

PDL was included at each amplifier with a value of $0.5$ dB. The
residual dispersion per span was either 30 ps/nm (dispersion-managed,
DM30) or absent (dispersion-uncompensated, DU). In any case, full
dispersion compensation was implemented at the receiver input after
propagation over $10$ or $20$ spans. Polarization-mode dispersion
was neglected since expected to be of minor concern \cite{Ivan_Amir}.

The link was simulated by the SSFM and compared with the prediction
of the PDL-EGN model. In the PDL-EGN matched filtering, zero-forcing
PDL equalization and carrier phase estimator (CPE) are implicit in
the model. On the other hand, in the SSFM case we implemented matched
filtering followed by a 1-tap zero-forcing equalizer, and  by a CPE
recovering the average phase induced by the fibers. 

We estimated the PDF of the received SNR by Monte Carlo simulations
over the PDL seeds, both with SSFM runs and with the PDL-EGN model.
In the SSFM case, we used 1000 different random PDL realizations.
For each realization, we varied the random state of polarization of
the channel lasers as well. The SSFM symmetric-step was chosen according
to the local-error criterion with a first step accumulating a FWM
phase of 20 rad \cite{Musetti}. The number of symbols was 65536,
sufficiently high to capture the largest walk-off among channels and
to have a negligible error from the Monte Carlo estimation \cite{Rossi}.
In the PDL-EGN, besides the Monte Carlo iterations over the PDL seeds,
we numerically solved the frequency integrals by another Monte Carlo
sampling.

\begin{figure}[tbh]
\begin{centering}
\includegraphics[width=0.8\linewidth]{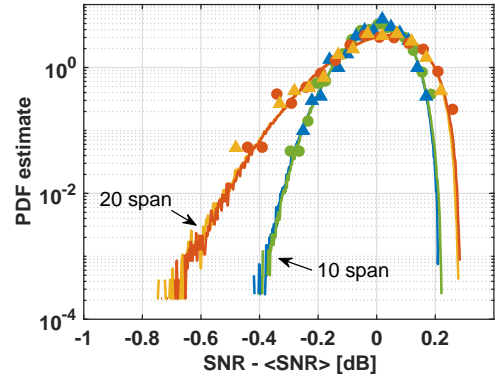}
\par\end{centering}
\caption{\label{fig:PDLGN}Estimate of the PDF of the SNR per polarization
by SSFM simulations (triangles for DM30, circles for DU) and the corresponding
(almost overlapped) PDL-GN PDFs (solid lines). }
\end{figure}

Fig.~\ref{fig:PDLGN} depicts the estimate of the PDF of the SNR
per polarization by SSFM (symbols) and the PDL-GN model (\ref{eq:SNRgn},
lines). We investigated two links lengths, 10 and 20 spans. In order
to compare the two links, we plotted the PDFs versus the SNR offset
from its mean. For each link, the mean SNR of SSFM simulations was
within 0.1 dB of that from the PDL-GN model.

We observe an excellent fit at both distances, thus confirming the
validity of the proposed model.  In particular, we note that dispersion
management does not affect the PDF shape, while it strongly impacts
the average value because of strongly different correlations of the
NLI among spans.

\begin{figure}[tbh]
\centering{}\includegraphics[width=0.8\linewidth]{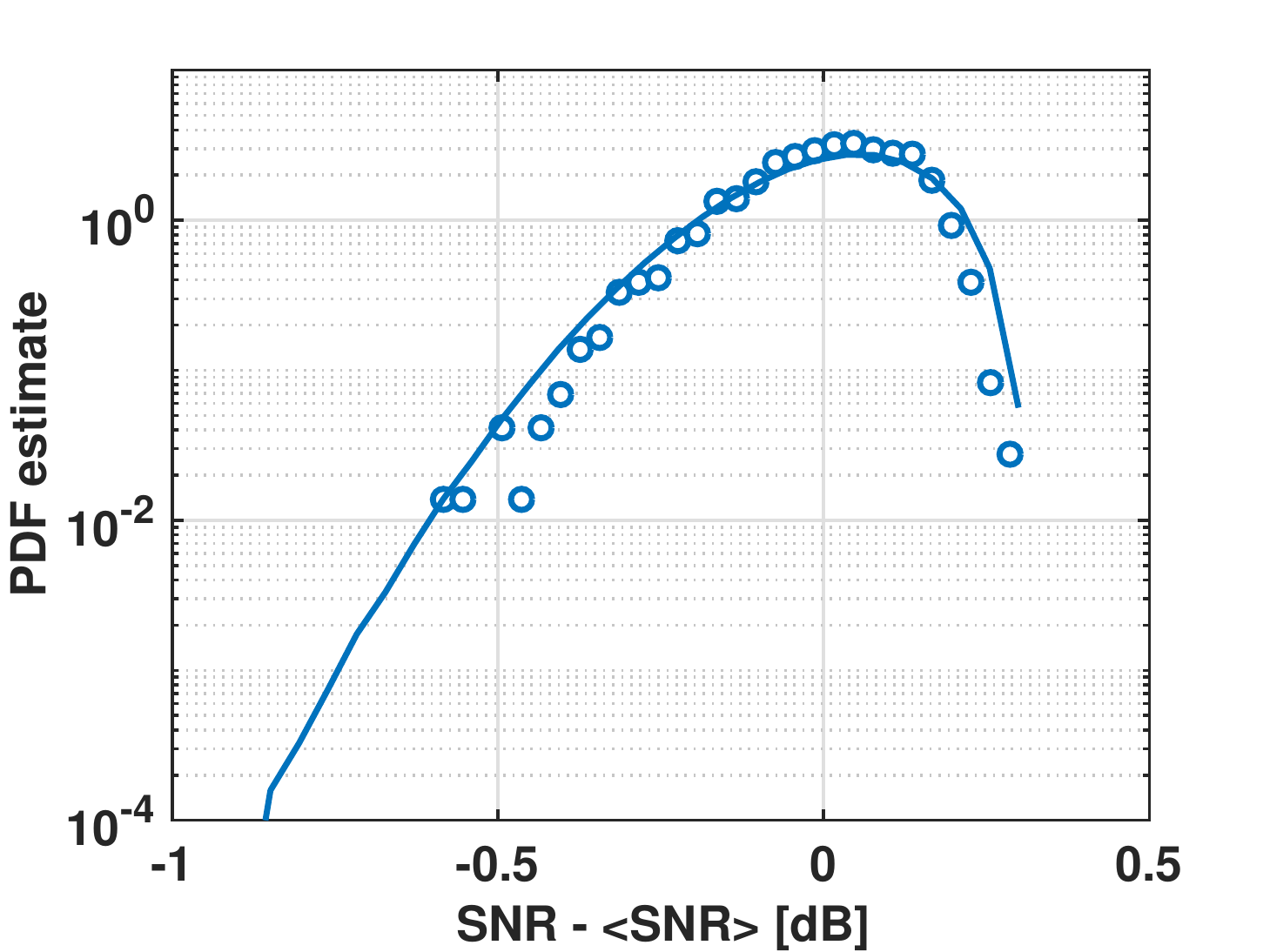}\caption{\label{fig:qam8}PDF estimate of the SNR per polarization by the PDL-EGN
(solid line) and SSFM (symbols). $32\times100$ km SMF link with star-8QAM-
modulation format, with different PDL between EDFA and ROADM.}
\end{figure}

\begin{figure}[tbh]
\begin{centering}
\includegraphics[width=0.8\linewidth]{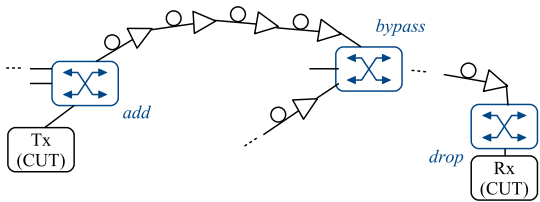}
\par\end{centering}
\caption{\label{fig:network}Network under investigation. 3200 km.}
\end{figure}

In a second test, we investigated a more realistic distribution of
PDL along the optical link. We thus focused on a $3200$ km network
scenario where the CUT, besides being added and dropped by ROADMs,
crosses ROADMs placed every 4 spans. Within each ROADM working in
bypass mode, two WSS were crossed. The PDL was 0.1 dB within EDFAs
and 0.4 dB within the WSS, respectively \cite{Cartledge_JLT19}. Fig.~\ref{fig:network}
sketches the link. 

The modulation format was PDM-star 8QAM, for a total of $21$ channels.
In this setup we included ASE, with a noise figure of 5 dB per amplifier.
It is worth noting that with ideal equalization the last drop impacts
equally ASE, NLI, and signal, hence with no implications on the statistics. 

The PDF estimated by the PDL-EGN and by SSFM at power 1 dBm maximizing
the SNR is reported in Fig.~\ref{fig:qam8}. Again, we observe a
good match between the two models. Even if not shown in the figure,
we observed a bias of 0.1 dB by the PDL-EGN model.

Having tested the validity of the model, we applied it to estimate
the outage probability, here defined as the probability that the SNR
falls below the threshold of 10.56 dB, corresponding to a Q-factor
of 6.5 dB. At such a threshold the achievable information rate of
a symbol-by-symbol star-8QAM detector in AWGN is 2.84 bits/symbol.
Since the loss to the nominal 3 bits/symbol carried out by the modulation
format is 0.16 bits/symbol, the transmission is expected to be feasible
with realistic forward-error correcting (FEC) codes. Moreover, since
the Shannon capacity at such threshold is 3.63 bits/symbol, the modulation
format is a good candidate for transmission over 3200 km. 

\begin{figure}[tbh]
\begin{centering}
\includegraphics[width=0.8\linewidth]{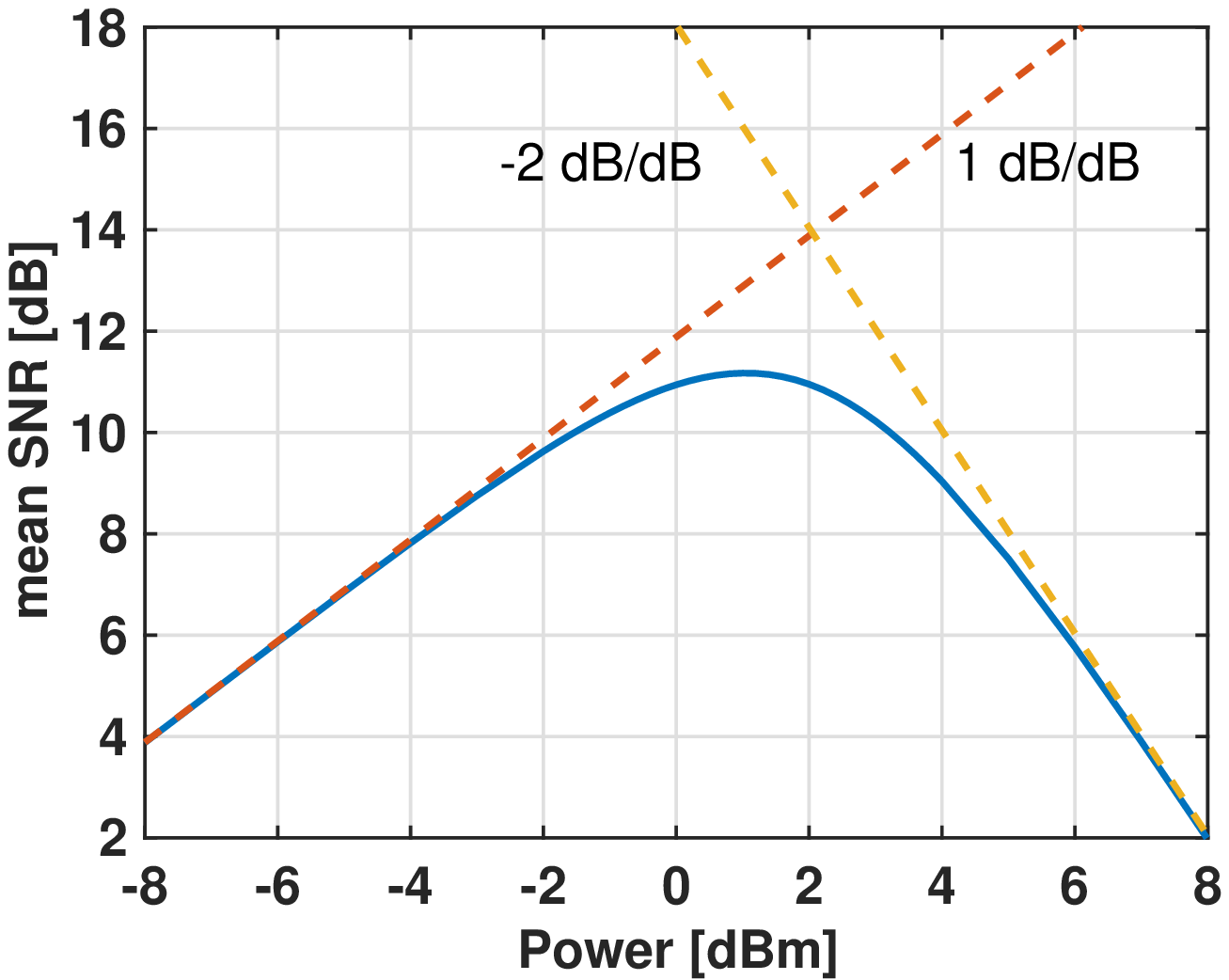}
\par\end{centering}
\begin{centering}
\includegraphics[width=0.8\linewidth]{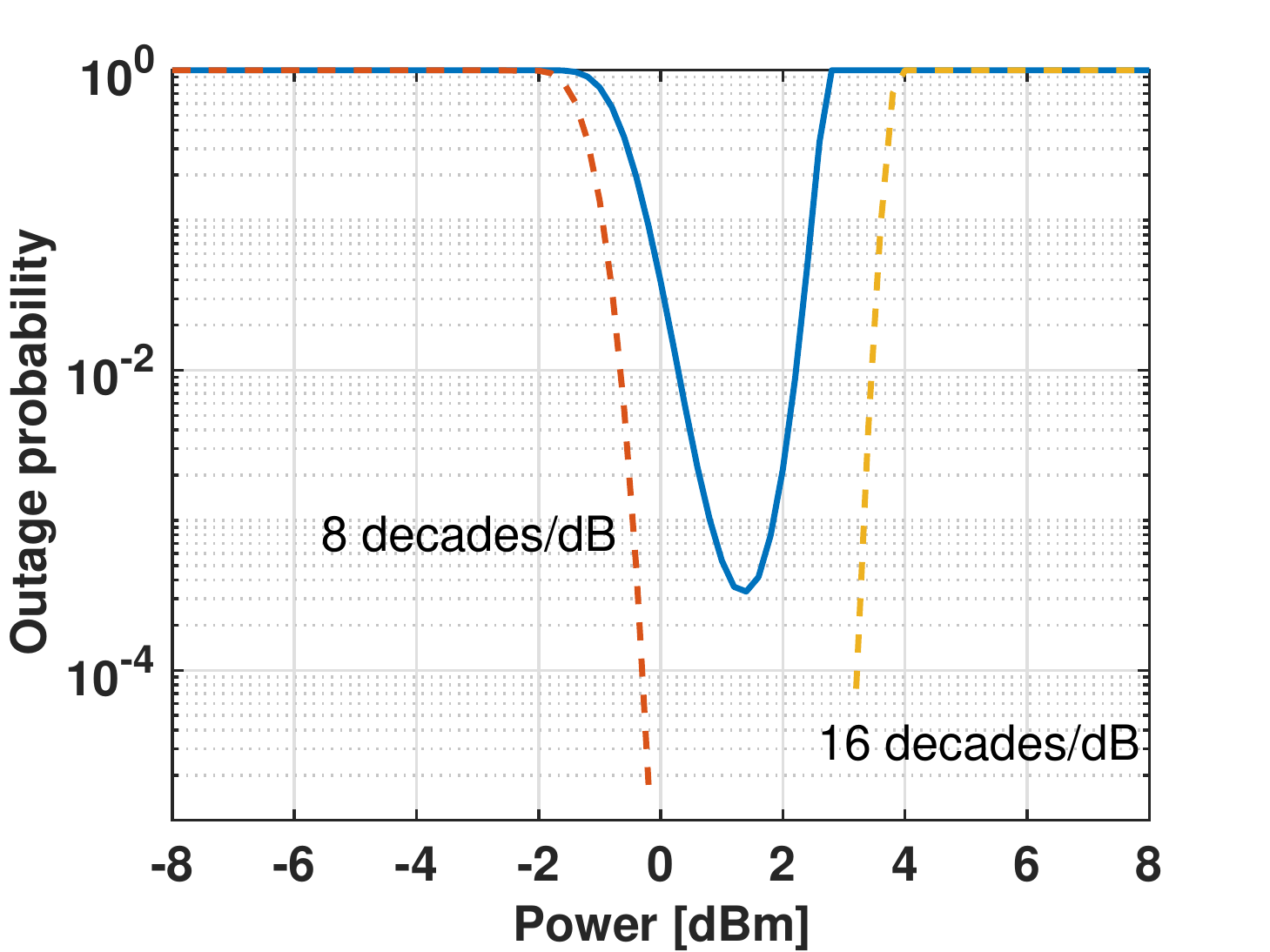}
\par\end{centering}
\caption{\label{fig:outage}Top: mean SNR per polarization vs. power. Bottom:
outage probability @ Q-factor = 6.5 dB. All curves with the proposed
model with only ASE or NLI (dashed lines) or with both ASE and NLI
(solid lines). Star-8QAM $32\times100$ km SMF link with different
PDL between EDFA and ROADM, see text.}
\end{figure}

Fig. \ref{fig:outage}(bottom) reports the outage probability versus
launched power per channel, and for reference Fig. \ref{fig:outage}(top)
also shows the mean SNR versus the same power. Several interesting
observations can be drawn from Fig.~\ref{fig:outage}. First, we
observe that the minimum outage probability is $3\cdot10^{-4}$, a
non-negligible value indicating the importance of including PDL in
link design. Second, the best power for the mean SNR does not coincide
with the best power for the outage probability, with a gap of 0.4
dB. This is strictly related to the nonlinear relation between the
outage probability and the SNR. Third, it is interesting to compare
the slope of the asymptotes in the ASE-dominated regime (linear regime)
and the NLI-dominated regime (nonlinear regime). In the mean SNR curve,
we observe a slope of +1 dB/dB in the linear regime and -2 dB/dB in
the nonlinear regime. A factor 2 in absolute terms between the two
slopes is well known and related to the scaling properties of the
ASE and the NLI variance \cite{Grellier}. Quite surprisingly, we
still observe a factor 2 between the slopes of the asymptotes on the
outage probability graphs. Such an observation can be very useful
for quickly scaling the outage probability with power. 

\begin{figure}[tbh]
\begin{centering}
\includegraphics[width=0.8\linewidth]{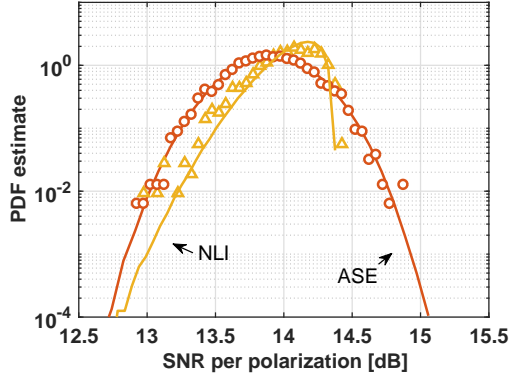}
\par\end{centering}
\caption{\label{fig:PDF_P2}Individual contribution of ASE and NLI to the PDF
by using the proposed PDL-EGN model (solid line) and SSFM simulations
(circles: ASE, triangles: NLI). Power $2$ dBm.}
\end{figure}

In Fig.~\ref{fig:PDF_P2} we compare the individual contributions
of ASE and NLI to SNR statistics. We set the power to $2$ dBm, i.e.,
where the two effects yield the same average variance. Symbols indicate
SSFM simulations. We observe that, in this setup, the PDF of the ASE-only
case is slightly larger than that of the NLI-only case, although the
average values are similar. Most important, the ASE-only and the NLI-only
PDFs have different shapes, hence the NLI cannot be treated as an
equivalent extra-ASE distributed along the link. 

\begin{figure}[tbh]
\begin{centering}
\includegraphics[width=0.8\linewidth]{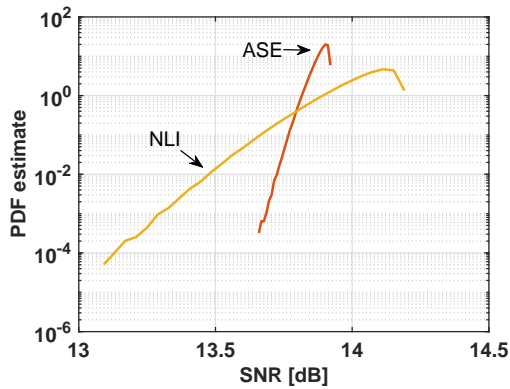}
\par\end{centering}
\caption{\label{fig:pdfQ_pdm}Estimate of the PDF of the PDM SNR usually adopted
in the literature by the proposed PDL-EGN model. Same setup of Fig.~\ref{fig:PDF_P2}. }
\end{figure}

It is worth noting that if instead of plotting the per-polarization
SNR PDF, we plot the PDF of the overall PDM SNR, as usually done in
the literature, we obtain a sharply different behavior, as depicted
in Fig.~\ref{fig:pdfQ_pdm} with the PDL-EGN model. The PDM SNR,
or simply SNR, is defined as $\text{SNR}=(S_{x}+S_{y})/(\sigma_{x}^{2}+\sigma_{y}^{2})$,
with $\sigma_{x,y}^{2}$ the variance of the noise under investigation.
The reason for the differences is related to the antithetic impact
of PDL on the $x$ and $y$ ASE variances, which is not manifested
by SPM- and XPM-like contributions that operate through a common scalar
nonlinear phase on both polarizations \cite{Rossi}.

It is interesting to compare the computational times of the PDL-EGN
model and the SSFM. As a reference, an SSFM simulation, with step
set-up as in \cite{Musetti}, took 1 day to run 125 PDL seeds on a
cluster using INTEL XEON E5- 2683v4 2.1GHz 32 cores central processing
units (CPU) with 128 GB of RAM and NVIDIA Tesla P100 graphics processing
unit (GPU). The same seeds have been simulated in a fraction of second
with the PDL-EGN, plus an overhead for the computation of the span
cross-correlations of the order of seconds for the PDL-GN and few
minutes for the PDL-EGN. Not surprisingly, with the EGN we were able
to simulate $10^{6}$ PDL seeds, while with SSFM only $10^{3}$.

\section{Conclusions\label{sec:Conclusions}}

We extended the EGN model to include PDL. The extension forced us
to rework the whole EGN theory, yielding a formally compact master
equation based on tensors able to span the time, frequency and polarization
axes. With the new model, it is possible to estimate the statistics
of the SNR in the nonlinear regime by exploring very rare events.
Our results show that the NLI interplay with PDL follows a different
behavior than the one experienced by ASE and thus needs a proper description.
Besides this aspect, probably the main advantage of the novel model
relies on its extremely fast computational time compared to standard
algorithms, such as the SSFM. For instance, we investigated 1000 different
optical links with random PDL in a fraction of second after a pre-processing
of the order of seconds to minutes. Such times are inaccessible to
SSFM simulations. The model can now be used to include PDL precisely
in the design of optical links, while today the problem is approximately
solved by allocating empirical margins to PDL. This way, it is possible
to perform the design optimizing the system outage probability instead
of the average performance. 

\appendices{}

\section{Nonlinear interference\label{App:Nonlinear-interference}}

In the absence of polarization-mode dispersion, the optical propagation
within an optical fiber is well described by the Manakov equation.
A first-order perturbative solution of the Manakov equation is \cite{Mecozzi,Poggiolini_GN,Dar_NLIN,Serena_EGN}:
\begin{equation}
\ket{A(z,t)}\simeq e^{{\cal L}z}\ket{A(0,t)}+\int_{0}^{z}e^{{\cal L}(z-\xi)}{\cal N}\left(e^{{\cal L}\xi}\ket{A(0,t)}\right)\text{d}\xi~.\label{eq:RP1-1}
\end{equation}
where the operator ${\cal L}$ accounts for linear effects while ${\cal N}$
for nonlinear effects, respectively. ${\cal L}$ is best defined in
the frequency domain by its Fourier transform $\mathscr{F}\left\{ e^{{\cal L}z}\right\} \triangleq e^{\vartheta(z,\omega)}$,
with $\vartheta$ accounting for attenuation and dispersion, see (\ref{eq:lineareff}).
${\cal N}$ is best described in the time domain $t$, i.e., 
\[
{\cal N}\triangleq-j\gamma\frac{8}{9}\braket{A(z,t)|A(z,t)}\ket{A(z,t)}~.
\]

The integral in (\ref{eq:RP1-1}) defines the NLI whose physical interpretation
is simple: the NLI is additive along distance, and its generic contribution
generated at coordinate $\xi$ depends on the unperturbed signal up
to that coordinate and experiences only linear effects up to the end
\cite{Serena_EGN}. If we factor out $e^{{\cal L}z}$ and apply zero-forcing
equalization, i.e., concatenation with the inverse $e^{-{\cal L}z}$,
the received signal $\ket{A_{\text{R}}}$ is:
\begin{equation}
\ket{A_{\text{R}}}\simeq\ket{A(0,t)}+\int_{0}^{z}e^{-{\cal L}\xi}{\cal N}\left(e^{{\cal L}\xi}\ket{A(0,t)}\right)\text{d}\xi~.\label{eq:Ar}
\end{equation}
which further suggests interpreting the NLI as an infinite summation
of echoes. 

The Fourier transform $\ket{\tilde{n}(\omega)}$ of the NLI in (\ref{eq:Ar})
is:
\begin{align*}
\ket{\tilde{n}(\omega)} & =-j\iint_{-\infty}^{\infty}\eta(\omega,\omega_{1},\omega_{2})\times\\
 & \braket{\tilde{A}(\omega+\omega_{1}+\omega_{2})|\tilde{A}(\omega+\omega_{2})}\ket{\tilde{A}(\omega+\omega_{1})}\frac{\text{d}\omega_{1}}{2\pi}\frac{\text{d}\omega_{2}}{2\pi}
\end{align*}
with the kernel of the optical link in $(0,z)$ given by:
\begin{multline}
\eta(\omega,\omega_{1},\omega_{2})\triangleq\\
\!\!\frac{8}{9}\gamma\int_{0}^{z}e^{\vartheta(\xi,\omega+\omega_{1})+\vartheta(\xi,\omega+\omega_{2})+\vartheta^{*}(\xi,\omega+\omega_{1}+\omega_{2})-\vartheta(\xi,\omega)}\text{d}\xi~.\label{eq:kernel}
\end{multline}
It is worth noting that for optical fibers with constant parameters
the integral in $\xi$ can be given in closed form \cite{Poggiolini_GN}.

Inserting a frequency-independent PDL in the model corresponds to
applying the substitutions $e^{{\cal L}\xi}\rightarrow e^{{\cal L}\xi}\mathbf{U}(z)$
and $e^{-{\cal L}\xi}\rightarrow e^{-{\cal L}\xi}\mathbf{U}^{-1}(z)$,
with $\mathbf{U}$ accounting for the cumulative PDL up to coordinate
$z$, as per (\ref{eq:lineareff}). With lumped PDL the matrix $\mathbf{U}$
is a staircase function in $z$, hence, it is convenient breaking
the integral in $z$ into a summation of integrals between consecutive
PDL elements, i.e., $\int_{0}^{z}\rightarrow\sum_{p}\int_{z_{p-1}}^{z_{p}}$.
With such substitutions we finally get:
\begin{multline}
\ket{\tilde{n}(\omega)}=-j\iint_{-\infty}^{\infty}\sum_{p=0}^{N-1}\eta_{p}(\omega,\omega_{1},\omega_{2})\times\\
\braket{\tilde{A}(\omega+\omega_{1}+\omega_{2})|\mathbf{P}(z_{p})|\tilde{A}(\omega+\omega_{2})}\ket{\tilde{A}(\omega+\omega_{1})}\frac{\text{d}\omega_{1}}{2\pi}\frac{\text{d}\omega_{2}}{2\pi}\label{eq:aw-2}
\end{multline}
where $\mathbf{P}(z_{p})\triangleq\mathbf{U}^{\dagger}(z_{p})\mathbf{U}(z_{p})$
and $\eta_{p}$ can be evaluated as per (\ref{eq:kernel}) but integrated
over $(z_{p},z_{p+1})$. If the source does not have PDL, we simply
set $\mathbf{U}(z_{0})=\mathbf{I}$ at $z_{0}=0$.

\section{Master theorem\label{App:Master-theorem}}

The key ingredient to perform a statistical analysis is the variance
of the NLI atom, i.e., $\mathbb{E}[|n_{\mathbf{i}}|^{2}]$:
\[
\mathbb{E}\left[n_{\mathbf{i}}n_{\mathbf{i}}^{*}\right]=\sum\mathbb{E}\left[a_{\mathbf{k}}^{*}a_{\mathbf{m}}a_{\mathbf{n}}a_{\mathbf{l}}a_{\mathbf{j}}^{*}a_{\mathbf{o}}^{*}\right]{\cal X}_{\mathbf{kmni}}{\cal X}_{\mathbf{ljoi}}^{*}~.
\]
However, part of the NLI is compensated by digital signal-processing
at the receiver. At the simplest level, a basic CPE removes the average
phase $\varphi$, which, in the perturbative framework, corresponds
to work with the following NLI:
\begin{equation}
n_{\mathbf{i}}^{\prime}=n_{\mathbf{i}}+j\varphi a_{\mathbf{i}}~.\label{eq:NLI_CPE}
\end{equation}
 The real target is thus $\mathbb{E}[|n_{\mathbf{i}}^{\prime}|^{2}]$,
which is given by the following:

\begin{theorempao}\label{thm:Master_space}Assume all the $a_{\mathbf{k}}$
to be complex zero-mean independent random variables with $n$-fold
rotational symmetry and $n\!\ge\!4$. Then:
\begin{multline}
\mathbb{E}\left[n_{\mathbf{i}}^{\prime}n_{\mathbf{i}}^{\prime*}\right]=\sum_{\mathbf{n}}\kappa_{3}^{(\mathbf{n})}\left|{\cal X}_{\mathbf{nnni}}\right|^{2}\\
+\sum_{\mathbf{k,n}}\kappa_{2}^{(\mathbf{k})}\kappa_{1}^{(\mathbf{n})}\Big(\left|{\cal X}_{\mathbf{kkni}}+{\cal X}_{\mathbf{knki}}\right|^{2}+\left|{\cal X}_{\mathbf{nkki}}\right|^{2}\Big)\\
+\sum_{\mathbf{k,m,n}}\kappa_{1}^{(\mathbf{k})}\kappa_{1}^{(\mathbf{n})}\kappa_{1}^{(\mathbf{m})}{\cal X}_{\mathbf{kmni}}\left({\cal X}_{\mathbf{kmni}}^{*}+{\cal X}_{\mathbf{knmi}}^{*}\right)\label{eq:Main_thm}
\end{multline}
with $\kappa_{n}^{(\mathbf{k})}$ the $n$-th order cumulant of data
symbols at channel use at time $k_{1}$, frequency $k_{2}$, and space
$k_{3}$.\hfill\IEEEQEDopen\end{theorempao}

Such a theorem is a generalization of the one provided in \cite{Serena_EGN}.
It is worth noting that the main difference is the broken degeneracy
between the inner indexes of the tensor. For instance, the last term
${\cal X}_{\mathbf{kmni}}^{*}+{\cal X}_{\mathbf{knmi}}^{*}$ is equal
to $2{\cal X}_{\mathbf{kmni}}^{*}$ in the scalar case according to
(\ref{eq:X_symmetry}). The basic intuition behind (\ref{eq:Main_thm})
is the following. Since the Kerr nonlinearity is cubic, the product
$n_{\mathbf{i}}n_{\mathbf{i}}^{*}$ depends on the product of six
atoms. Only combinations with an equal number of conjugate/non-conjugate
pairs are non-zero, as depicted in Fig.~\ref{fig:combinations}.
Each combination is weighted by the corresponding symbol-cumulant
\cite{Serena_EGN}. Please note that terms labeled with - in Fig.~\ref{fig:combinations}
correspond to the terms removed by the CPE in (\ref{eq:NLI_CPE})
\cite{serena_GN_JLT}. 

The indexing of the valid combinations yields the terms in (\ref{eq:Main_thm}).
Hence, for instance, the PDL-GN model reduces to the last triple summation
in (\ref{eq:Main_thm}). 

\begin{figure}[tbh]
\begin{centering}
\begin{tabular}{c|c|c||c|c|c|c}
\hline 
$\mathbf{k}^{*}$ & $\mathbf{m}$ & $\mathbf{n}$ & $\mathbf{l}$ & $\mathbf{j}^{*}$ & $\mathbf{o}^{*}$ & \tabularnewline
\hline 
\hline 
$\bullet$ & $\bullet$ & $\bullet$ & $\bullet$ & $\bullet$ & $\bullet$ & Q6\tabularnewline
\hline 
\hline 
$\bullet$ & $\bullet$ & $\circ$ & $\bullet$ & $\bullet$ & $\circ$ & F4\tabularnewline
\hline 
$\bullet$ & $\bullet$ & $\circ$ & $\bullet$ & $\circ$ & $\bullet$ & F4\tabularnewline
\hline 
$\bullet$ & $\circ$ & $\bullet$ & $\bullet$ & $\bullet$ & $\circ$ & F4\tabularnewline
\hline 
$\bullet$ & $\circ$ & $\bullet$ & $\bullet$ & $\circ$ & $\bullet$ & F4\tabularnewline
\hline 
$\bullet$ & $\bullet$ & $\bullet$ & $\circ$ & $\bullet$ & $\circ$ & \emph{-}\tabularnewline
\hline 
$\bullet$ & $\bullet$ & $\bullet$ & $\circ$ & $\circ$ & $\bullet$ & \emph{-}\tabularnewline
\hline 
$\circ$ & $\bullet$ & $\circ$ & $\bullet$ & $\bullet$ & $\bullet$ & \emph{-}\tabularnewline
\hline 
$\circ$ & $\circ$ & $\bullet$ & $\bullet$ & $\bullet$ & $\bullet$ & \emph{-}\tabularnewline
\hline 
$\circ$ & $\bullet$ & $\bullet$ & $\circ$ & $\bullet$ & $\bullet$ & Q4\tabularnewline
\hline 
\hline 
$\bullet$ & $\bullet$ & $\circ$ & $\vartriangle$ & $\vartriangle$ & $\circ$ & \emph{-}\tabularnewline
\hline 
$\bullet$ & $\bullet$ & $\circ$ & $\vartriangle$ & $\circ$ & $\vartriangle$ & \emph{-}\tabularnewline
\hline 
$\bullet$ & $\circ$ & $\bullet$ & $\vartriangle$ & $\vartriangle$ & $\circ$ & \emph{-}\tabularnewline
\hline 
$\bullet$ & $\circ$ & $\bullet$ & $\vartriangle$ & $\circ$ & $\vartriangle$ & \emph{-}\tabularnewline
\hline 
$\bullet$ & $\vartriangle$ & $\circ$ & $\bullet$ & $\vartriangle$ & $\circ$ & GN\tabularnewline
\hline 
$\bullet$ & $\vartriangle$ & $\circ$ & $\bullet$ & $\circ$ & $\vartriangle$ & GN\tabularnewline
\hline 
\end{tabular}
\par\end{centering}
\caption{\label{fig:combinations}Valid combinations yielding non-zero $\mathbb{E}\left[a_{\mathbf{k}}^{*}a_{\mathbf{m}}a_{\mathbf{n}}a_{\mathbf{l}}a_{\mathbf{j}}^{*}a_{\mathbf{o}}^{*}\right]$.
For instance, the second row indicates the combination $(\mathbf{k}\!=\!\mathbf{m}\!=\!\mathbf{l}\!=\!\mathbf{j})\protect\ne(\mathbf{n}\!=\!\mathbf{o})$.
The labels indicate sixth-order noise (Q6), two types of fourth-order
noise (F4 and Q4) and second-order noise, usually called GN contribution.
A (-) indicates a phase contribution that is removed by the CPE.}
\end{figure}

Although the master theorem reduces the number of summations, the
final result (\ref{eq:Main_thm}) still depends on an infinite summation
over the discrete-time index. However, in the special case of sinc
pulses, such a summation can be dropped, with significant simplifications,
as already observed in \cite{Mecozzi,Dar_NLIN}, thanks to the Poisson
summation formula:
\begin{equation}
\sum_{k=-\infty}^{\infty}e^{jk\omega T}=\frac{2\pi}{T}\sum_{k=-\infty}^{\infty}\delta\left(\omega-\frac{2\pi k}{T}\right)\label{eq:Poisson_formula}
\end{equation}
and the finite bandwidth of the pulses that can interact with only
one Dirac's delta in such a summation. Instead of entering the fine
mathematical details of the proof, the next theorem provides a short-rule
for the simplifications:

\begin{theorempao}With sinc pulses $p_{h}(t)=\text{sinc}(t/T)$,
$\forall h$, all summations in (\ref{eq:Main_thm}) over \emph{temporal}
indexes can be dropped. For each drop, an integral in the corresponding
tensor-product ${\cal X}_{\mathbf{kmni}}{\cal X}_{\mathbf{ljoi}}^{*}$
can be dropped as well. The dropped integral can be identified by
solving a linear system obtained by equating the arguments $\psi$
of the $\ket{\tilde{G}_{\mathbf{k}}(\psi)}$ involved in the product
owing to the same atom.\hfill\IEEEQEDopen\end{theorempao}

Such a result is best explained by an example. Let us focus on the
particular GN-term ${\cal G}={\cal X}_{\mathbf{kmni}}{\cal X}_{\mathbf{kmni}}^{*}$
of (\ref{eq:Main_thm}) in the scalar case. This term can be explicitly
expanded following (\ref{eq:tensor_S}) as:
\begin{align*}
{\cal G} & =\sum_{t,r=0}^{N-1}\iiint\!\!\!\iiint_{-\infty}^{\infty}\eta_{t}(\omega,\omega_{1},\omega_{2})\eta_{r}^{*}(\mu,\mu_{1},\mu_{2})\\
 & \quad\times\tilde{G}_{\mathbf{k}}^{*}(\omega+\omega_{1}+\omega_{2})\tilde{G}_{\mathbf{m}}(\omega+\omega_{2})\tilde{G}_{\mathbf{i}}^{*}(\omega)\tilde{G}_{\mathbf{n}}(\omega+\omega_{1})\\
 & \quad\times\tilde{G}_{\mathbf{k}}(\mu+\mu_{1}+\mu_{2})\tilde{G}_{\mathbf{m}}^{*}(\mu+\mu_{2})\tilde{G}_{\mathbf{i}}(\mu)\tilde{G}_{\mathbf{n}}^{*}(\mu+\mu_{1})\\
 & \frac{\text{d}\omega_{1}}{2\pi}\frac{\text{d}\omega_{2}}{2\pi}\frac{\text{d}\omega}{2\pi}\frac{\text{d}\mu_{1}}{2\pi}\frac{\text{d}\mu_{2}}{2\pi}\frac{\text{d}\mu}{2\pi}~.
\end{align*}
Equating the arguments of equal atoms yields the following linear
system:
\[
\begin{cases}
\mathbf{k} & \rightarrow\omega+\omega_{1}+\omega_{2}=\mu_{1}+\mu_{2}+\mu\\
\mathbf{m} & \rightarrow\omega+\omega_{2}=\mu+\mu_{2}\\
\mathbf{n} & \rightarrow\omega+\omega_{1}=\mu+\mu_{1}
\end{cases}
\]
whose solution is $\omega_{1}=\mu_{1},~\omega_{2}=\mu_{2},~\omega=\mu$.
We can thus drop three integrals, for instance the ones with $(\mu,\mu_{1},\mu_{2})$,
by using the previous substitution.

\end{document}